%% file: JRWRTC2016.tex
\documentclass{sig-alternate}

\usepackage{amsmath, amsfonts, amssymb}
\usepackage{stmaryrd}
\usepackage{multirow}
\usepackage{hyperref}
\usepackage{tikz}
\usetikzlibrary{shapes,decorations,positioning}

\newtheorem{definition}{Definition}

\emergencystretch=\maxdimen
\makeatletter
\def\@copyrightspace{\relax}
\makeatother

\begin{document}
\title{Model Checking of Cache for WCET Analysis Refinement
\thanks{This work was partially supported by the
\href{http://erc.europa.eu/}{European Research Council} under the
European Union's Seventh Framework Programme (FP/2007-2013) / ERC Grant
Agreement nr. 306595 \href{http://stator.imag.fr}{``STATOR''.}}
\conferenceinfo{JRWRTC '16}{October 19--21, 2016, Brest, FRANCE}
}

\numberofauthors{3}

\author{
	\alignauthor
	Valentin Touzeau\\
	\affaddr{Univ. Grenoble Alpes\\}
	\affaddr{CNRS, VERIMAG, F-38000\\}
	\affaddr{Grenoble, France\\}
	\email{Valentin.Touzeau@imag.fr}
	\alignauthor
	Claire Ma{\"i}za\\
	\affaddr{Univ. Grenoble Alpes\\}
	\affaddr{CNRS, VERIMAG, F-38000\\}
	\affaddr{Grenoble, France\\}
	\email{Claire.Maiza@imag.fr}
	\alignauthor
	David Monniaux\\
	\affaddr{Univ. Grenoble Alpes\\}
	\affaddr{CNRS, VERIMAG, F-38000\\}
	\affaddr{Grenoble, France\\}
	\email{David.Monniaux@imag.fr}
}

\maketitle

\begin{abstract}
	On real-time systems running under timing constraints, scheduling can be performed when one is aware of the worst case execution time (WCET) of tasks. Usually, the WCET of a task is unknown and schedulers make use of safe over-approximations given by static WCET analysis. To reduce the over-approximation, WCET analysis has to gain information about the underlying hardware behavior, such as pipelines and caches. In this paper, we focus on the cache analysis, which classifies memory accesses as hits/misses according to the set of possible cache states. We propose to refine the results of classical cache analysis using a model checker, introducing a new cache model for the least recently used (LRU) policy.
\end{abstract}


\section{Introduction}


On critical systems, one should be able to guarantee that each task will meet its deadline. This strong constraint can be satisfied when the scheduler has bounds on every tasks' execution time. The aim of a WCET analysis is to compute such safe bounds statically. In order to provide a satisfiable bound, the WCET analysis needs to model the execution of instructions at the hardware level. However, to avoid the huge latency of main memory access, one can copy parts of the main memory into small but fast memories called caches. In order to retrieve precise bounds on the execution time of instructions, it is thus mandatory to know which instructions are in the cache and which are not. In this paper we focus on instruction caches, ie. we aim at knowing whether instructions of the program are in the cache when they are executed.

For efficiency reasons, the main memory is partitioned into fixed size blocks. To avoid repeated accesses to the same block, they are temporary copied into the cache when requested by the CPU. This way, they can be retrieved faster on the next access, without requesting the main memory again. Moreover, to speed up the retrieval of blocks from the cache, caches are usually partitioned into sets of equal sizes. A memory block can only be stored in one set that is uniquely determined by the address of the block. Thus, when searching a block in the cache, it is looked for in only one set.
Since the main memory is bigger than the cache, it may happen that a set is already full when trying to store a new block in it. In this case, one block of the set has to be evicted in order to store the new one. This choice does not depend on the content of the other sets and is done according to the cache replacement policy. In our case, we focus on the Least Recently Used (LRU) policy (for other policies, refer to~\cite{DBLP:journals/lites/LvGRW016}). A cache set using the LRU policy can be represented as a queue containing blocks ordered from the most recently accessed (or used) to the least recently accessed. When the CPU requests a block that is not in the cache (cache miss), it is stored at the beginning of the queue (it becomes the most recently used block) and the last block (the least recently used) is evicted. On the other hand, when the requested block is already in the cache, it is moved to the beginning of the queue, delaying its eviction. The position of a block in the queue is called the age of the block: the youngest block is the most recently used and the oldest is the least recently used.

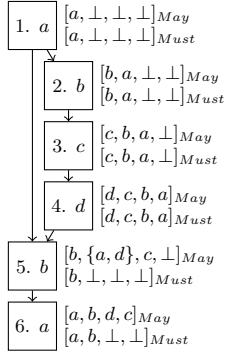
\begin{figure}[!htb]
	\centering
	\input{unknownExample.tex}
	\caption{\label{UnknownAccessExample}Example of access classified as "unknown"}
\end{figure}

The aim of a cache analysis is to provide a classification of memory accesses as "cache hit", "cache miss" or "unknown" (not always a hit, and not always a miss) to be used as part of the WCET analysis. This classification is usually established by abstract interpretations called "May Analysis" and "Must Analysis" that respectively compute a lower and upper bound of every block's age. For more details about these analysis, refer to~\cite{DBLP:books/daglib/0091605}. \emph{Must} analysis is used to predict the cache hits (if a block must be in the cache when accessing it, access is a hit), whereas \emph{may} analysis is used to predict the misses (if a block may not be in the cache when accessing it, access is a miss). However, if a block is in the \emph{may} cache but not in the \emph{must} cache, it is classified as unknown. This can happen because this access is sometimes a miss and sometimes a hit, or because the abstract interpretation is too coarse. An example is given on Figure~\ref{UnknownAccessExample}. Program states (basic blocks) are on the left, whereas abstract cache states (\emph{may} and \emph{must}) at the exit of basic blocks are on the right. For simplicity, every basic block is stored in exactly one memory block. For example, at the exit of block~5, the minimum age of blocks $a$, $b$, $c$ and $d$ computed by the \emph{may} analysis are respectively 1, 0, 2 and 1. At block~6, $a$ is accessed and is in the cache (because there are only 4 different blocks, and they all fit together in the cache), thus it should be classified as a hit. However, it is classified as "unknown" by the \emph{may} and \emph{must} analysis because of an over-approximation performed by the \emph{must} analysis. Indeed, at entry of block~5, the \emph{must} cache is $[\bot, \bot, \bot, a]$ because $a$ is the only block that must be accessed, and $b$, $c$ and $d$ may be accessed since the last access to $a$.
An other possibility to classify memory access as hit or miss is to use a model checker. Both the program and the cache are encoded as a transition system. Then, the question of the existence in the cache of a given block at a given program point is encoded as a logical formula. Both the formula to check and the transition system are provided to the model checker, that classifies the block as ``in the cache'', ``out of the cache'' or ``unknown''. Since the model deals with every reachable program/cache states separately, model checking is usually more precise than abstract interpretation. However, it is also much slower and often requires more memory during the analysis.

To avoid the precision loss of abstract interpretation without performing a heavy analysis using a model checker, we propose to mix both analysis. We first perform the classical \emph{may}/\emph{must} analysis by abstract interpretation, and then refine the classification using a model checker. Thus, we only use the model checker to classify blocks that were classified as "unknown" by the abstract interpretation. Moreover, we introduce a new abstract model that can be used by the model checker to efficiently represent LRU cache states.

\section{WCET Analysis}

This section gives some basic notions about WCET analysis and explain the link with cache analysis.




Usually, WCET analysis are performed by following steps: First, a control flow graph is retrieved from the binary code under analysis by grouping instructions into basic blocks (sequences of instructions with only one entry point and one exit point). Then, the WCET of the program is computed by bounding the execution time of every basic block and finding the ``longest'' path inside the CFG.

The computation of basic blocks execution time is done by micro-architectural analysis that models pipelines and caches. Independently, several other analysis like value analysis and loop bound analysis are performed to provide information about the semantics to the WCET analyzer.

At hardware level, the uncertainty about execution times of instructions comes from pipelines (which can start executing an instruction before the previous one is finished) and caches (which avoid costly main memory accesses). The aim of a cache analysis is to classify memory accesses as cache hit or cache miss. Since an access to the main memory can be 100 times slower than an access to a cache, it is mandatory to classify memory access as hit/miss in order to provide accurate bounds on WCET estimations. Moreover, when a memory access is not classified as a hit or a miss, the WCET analysis must treat the both cases~\cite{Cullmann_predictabilityconsiderations}. Thus, it can make the analysis slower and increase the pessimism.

To avoid this precision loss, we aim at refining the classification of unknown block using a model checker.






\section{Related Work}
Some previous works use a model checker for performing WCET analysis of programs. The approach of Lv, Yi, Guan and Yu~\cite{DBLP:conf/rtss/LvYGY10} is the following. Using abstract interpretation, they classify memory access as "cache misses", "cache hits" or "unknown". Every memory access is translated in a timed automaton that describe the access to the memory bus by introducing some delay. Finally, these automata are connected together according to the CFG and the model checker explores the transition system, computing a WCET estimation. This approach allows them to perform WCET analysis of multicore systems. Although they are using a model checker in a WCET analysis, our approach is different and complementary. Indeed, they are using the model checker to refine the timing analysis itself, and not the classification of memory accesses as we do. Therefore, our refinement of the cache content can be added to the first step of their analysis to retrieve better bounds. The work of Chattopadhyay and Roychoudhury~\cite{DBLP:conf/rtss/ChattopadhyayR11} is closer to our approach. They use the model checker to analyze behavior of caches shared by several cores. Moreover, by instrumenting the program at source code level they can take the program semantics into account and do not treat infeasible path, make the analysis more precise. Since they are performing a \emph{may}/\emph{must} analysis as a first step, we believe our analysis can be used to refine this first step, before taking shared caches into account.

\section{Our Abstract Model}

To perform the cache analysis using a model checker, we have to provide models both for the program and for the cache. We use these models to answer the following questions: ``At a given program point, is a given block in the cache whatever the path to reach this point is ?'' (classify as hit) and ``Is the given block never in the cache at the given program point, whatever the path used to reach this point is ?'' (classify as miss).

%

To model the cache, we use an abstraction of the real cache state to avoid the state space explosion problem one can meet when using a model checker. In the following formal description of our model, we adapt the notations from Jan Reineke's PhD~\cite{DBLP:phd/de/Reineke2009}:
\begin{definition} Cache size
	\begin{flalign*}
		& k \in \mathbb{N} \text{ is the size of the cache set (in blocks)}&
	\end{flalign*}
\end{definition}
\begin{definition} Set of memory blocks
	\begin{flalign*}
		& M\text{ represents the set of memory blocks.} &\\
		& M_{\bot} = M \cup \{\bot\}\text{ includes the empty line.}&
	\end{flalign*}
\end{definition}
\begin{definition} Set of Cache States
	\begin{flalign*}
		&C^{LRU} \subset M_\bot^k\text{ symbolizes the set of reachable cache states}& \\
		& [b_1, b_2, \dots, b_k] \in C^{LRU} \text{ represents a reachable state}& \\
		& b_1 \text{ is the least recently used block}
	\end{flalign*}
\end{definition}

In addition of these notation, to define our abstract model, we introduce the following notations: $\mathcal{A}$, represents the set of abstract cache states, $\varepsilon_a \in \mathcal{A}$, represents cache states that does not contain $a$, and $[S]_a \in \mathcal{A}$, represents cache states that contains $a$ and some other blocks younger than $a$ (forming the set $S$), where $a \in M$ is a memory block.
 
\begin{definition} Set of Abstract Cache States
	\begin{flalign*}
		& \mathcal{A} = \mathcal{P}(C^{LRU})\text{ is the power set of reachable cache states.}&
	\end{flalign*}
\end{definition}

\begin{definition} Abstract Cache States
	\begin{flalign*}
		&\varepsilon_a = \left\{ [b_1,...,b_k] \in C^{LRU}, \forall i \in \llbracket 1,k \rrbracket, b_i \neq a \right\} \in \mathcal{A}&
	\end{flalign*}
	\begin{multline*}
		[S]_a = \left\{ [b_1,...,b_k] \in C^{LRU},\right.\\
		\left.\vphantom{C^{LRU}} \exists i \in \llbracket 1,k \rrbracket, b_i = a \wedge \left( b_j \in S \Leftrightarrow j < i \right) \right\} \in \mathcal{A}
	\end{multline*}
\end{definition}

\begin{table*}
	\scriptsize
	\centering
	\begin{tabular}{|c|c|c|c|c|c|c|c|c|}
		\hline
		\multirow{2}{*}{Program} & \multirow{2}{*}{Size} & \multicolumn{2}{|c|}{4 ways} & \multicolumn{2}{|c|}{8 ways} & \multicolumn{2}{|c|}{16 ways}\\
		\cline{3-8}
				&        & Un		& Nc		& Un		& Nc		& Un		& Nc		\\
		\hline
		recursion	& 26	& 34.6\%	& 11.1\%	& 53.8\%	& 7.1\%		& 53.8\%	& 21.4\%	\\
		fac		& 26	& 34.6\%	& 11.1\%	& 46.1\%	& 8.3\%		& 46.1\%	& 41.6\%	\\
		binarysearch	& 48	& 12.5\%	& 0\%		& 56.2\%	& 29.6\%	& 52.0\%	& 12.0\%	\\
		prime		& 57	& 10.5\%	& 0\%		& 29.8\%	& 35.2\%	& 57.8\%	& 18.1\%	\\
		insertsort	& 58	& 23.7\%	& 28.5\%	& 28.8\%	& 11.7\%	& 55.9\%	& 9.0\%		\\
		bsort		& 62	& 30.6\%	& 57.8\%	& 53.2\%	& 6.0\%		& 62.9\%	& 5.1\%		\\
		duff		& 64	& 10.9\%	& 0\%		& 37.5\%	& 12.5\%	& 37.5\%	& 12.5\%	\\
		countnegative	& 65	& 21.5\%	& 21.4\%	& 43.0\%	& 21.4\%	& 52.3\%	& 20.5\%	\\
		st		& 137	& 14.5\%	& 30.0\%	& 43.7\%	& 13.3\%	& 69.3\%	& 5.2\%		\\
		ludcmp		& 179	& 11.1\%	& 5.0\%		& 39.6\%	& 15.4\%	& 67.5\%	& 4.1\%		\\
		minver		& 265	& 20.7\%	& 29.0\%	& 44.1\%	& 12.8\%	& 63.0\%	& 10.7\%	\\
		statemate	& 582	& 7.5\%	& 2.2\%		& 7.9\%		& 4.3\%		& 8.2\%		& 2.0\%		\\
		\hline
	\end{tabular}

	\caption{\label{results} Precision of \emph{May}/\emph{Must} analysis and Model Checker}
\end{table*}

The idea behind the abstract model we define below is to track only one block (noted $a$). Indeed, to know whether a block is in a LRU cache, you only have to count the number of accesses made to pairwise different blocks since the last access to it. In other words, you do not have to know the age of others blocks, you are only interested in knowing if they are younger than the block you are tracking. Therefore, we group together cache states that have the same set of blocks younger than the block we want to track.

To every cache state $p$, we associate an abstract state $\alpha_a(p)$ which consists of the set of values younger than $a$ in the cache or a special value in the case where $a$ is not in the cache.
\begin{definition} Abstraction of Cache States
	\begin{flalign*} 
		&\alpha_a : C^{LRU} \rightarrow \mathcal{A} \\
		&\alpha_a \left( [b_1,...,b_k] \right) =
		\begin{cases}
			\varepsilon_a & \text{if } \forall i \in \llbracket 1,k\rrbracket, b_i \neq a \\
			[ \{ b_1,...,b_{i-1} \} ]_a & \text{if } \exists i \in \llbracket 1,k \rrbracket, b_i = a
		\end{cases}&
	\end{flalign*}
\end{definition}

For example, when tracking block $a$, the abstract cache state associated to the exit of block~1 of Figure~\ref{UnknownAccessExample} is $[\{\}]_a$, symbolizing that $a$ is the least recently used block at this point (the set of younger blocks is empty). At the exit of block~4, the abstract cache state is $[\{b,c,d\}]_a$.

Additionally, we define the partial function $update_{LRU(k)}^a$, which models the effect of accessing a block on an abstract state. When the abstract cache state does not contain $a$ (i.e. is equal to $\varepsilon_a$), it remains unchanged until an access to $a$ is made. When $a$ is accessed, every new block access appears into the set $S$. When the cardinal of $S$ reaches $k-1$ ($a$ is the least recently used block), a new access to a different block evicts $a$ (and new abstract cache state is reset to $\varepsilon_a$). If an access to $a$ is done in the meantime, the set $S$ of younger block is reset to $\{\}$.
\begin{definition} Abstract State Update
	\begin{flalign*}
		& update_{LRU(k)}^a : \mathcal{A} \times M \rightarrow \mathcal{A} \\
		& update_{LRU(k)}^a(\varepsilon_a, c) = 
		\begin{cases}
 			[\{\}]_a & \text{if } a = c \\
 			\varepsilon_a & \text{if } a \neq c
		\end{cases}&
	\end{flalign*}
	\begin{multline*}
 		update_{LRU(k)}^a([S]_a, c) = \\
		\begin{cases}
 			[\{\}]_a & \text{if } c = a\\
			[S]_a & \text{if } c \neq a \text{ and } c \in S \\
			[S \cup \{c\}]_a & \text{if } c \neq a \text{ and } c \notin S \text{ and } |S| < k-1 \\
 			\varepsilon_a & \text{if } c \neq a \text{ and } c \notin S \text{ and } |S| = k-1
		\end{cases}
	\end{multline*}
\end{definition}

Considering the example of Figure~\ref{UnknownAccessExample}, the model checker associates two different abstract states to block~5 depending on the incoming flow from block~1 or block~4. These states are respectively $[\{\}]_a$ and $[\{b,c,d\}]_a$. Thus, applying the update function for treating the access made to $b$ in block~5, we obtain $[\{b\}]_a$ and $[\{b,c,d\}]_a$. Therefore, we know that $a$ is not evicted from the cache by block~5 and access made to $a$ in block~6 is not classified as unknown anymore but as a hit.

The second part of our model is the model of the program. Since we focus on instruction caches, the model we use for the program is a graph obtained from the CFG by splitting basic blocks (when needed) into blocks of the size of a memory block. Thus, a path in the model represents the sequence of memory access that the instruction cache handles during the execution of the program. However, because we only track one memory block at a time, it is also possible to simplify the control flow graph according to this block. Indeed, one can slice the CFG according to the cache set associated to the block we want to track, removing every memory access to an other cache set. Moreover, we can remove from the obtained graph every node that is not an access to $a$ and that does not contribute to $a$ eviction. Thus, it appears that every node that does not contain $a$ in their entry \emph{may} cache can be removed.

\begin{figure}[!htb]
	\centering
	\input{simplified_graph}
	\caption{\label{SimplifiedCFG}Simplifying CFG according to access to $a$}
\end{figure}
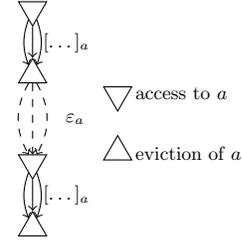

This simplification of the CFG is illustrated on Figure~\ref{SimplifiedCFG}. Plain arrows represent program flow potentially manipulating block $a$, whereas dashed arrows represent flow that does not and that can be simplified in only one arrow. At some point after an access to block $a$, we can be sure that $a$ is not in the cache anymore. Therefore, it is possible to remove all nodes (dashed arrows) from this point until the next access to $a$.



\section{Implementation / Experiments}

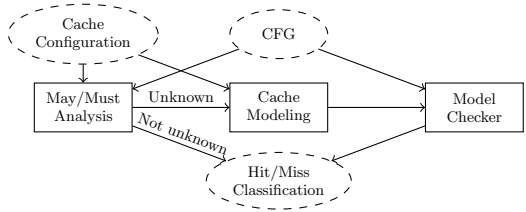
\begin{figure}[!htb]
	\centering
	\input{workflow}
	\caption{\label{workflow}Workflow of our prototype}
\end{figure}

This section describe the prototype we build and the experiments we made to valid our proof of concept. The workflow of our analysis is illustrated on Figure~\ref{workflow}.

Our implementation does not use directly the binary code to analyze but runs on the LLVM bytecode representation of it. We first build the CFG of the program from the bytecode using the LLVM framework. Since the LLVM bytecode does not affect any address to instructions, we have to provide a mapping of instructions to the main memory. For our prototype, we assume that every instruction has the same size in memory. Thus, memory blocks contain a fixed number of instructions and can be obtain by splitting basic blocks of the CFG into fixed size blocks. Using this mapping and the CFG, our prototype performs a \emph{may}/\emph{must} analysis of the program. For every block access classified as unknown, we build an abstract model of the cache and provide it to the model checker together with the CFG (simplified as explain above). It would be possible to use real addresses from binary code and a correspondence between LLVM bytecode and binary code, as done in~\cite{DBLP:conf/lctrts/HenryAMM14}, but this requires significant engineering and falls outside of the scope of this experiment.

We experiment our prototype with benchmarks of the TacleBench\footnote{http://www.tacle.eu/index.php/activities/taclebench}. Table~\ref{results} contains the results of our experiments. Size of program is given in number of memory block. We run our experiments with caches of only one cache set, with different sizes: 4, 8 or 16 ways. For every experiment, we measure both the amount of accesses classified as ``unknown'' by the \emph{may}/\emph{must} analysis (column ``Un'') and the amount of accesses newly classified as ``always in the cache'' or ``always out of the cache'' among the accesses left ``unknown'' by the \emph{may}/\emph{must} analysis (column Nc). During these experiments, our analysis classifies up to 57.8\% of the accesses left unclassified by the abstract interpretation analysis.


\section{Conclusion}

We proposed to refine classical cache analysis by using a model checker. To avoid the common problem of state space explosion meet when dealing with model checking, we introduce a new abstract cache model. This model allows us to compute the exact age of a memory block along an execution path of the program. Thus, we can select the memory block we want to refine. Moreover, it allows us to simplify the program model too, by removing some nodes useless to the refinement. Finally, we implement a prototype and test it on a benchmark. Our experiments shows that our approach is able to refine up to 60\% of the memory access classified as unknown by the abstract interpretation.

Our prototype runs on LLVM bytecode, and use an unrealistic memory mapping. As future work, we aim at implementing an analyzer that runs on the binary code. To finally validate our approach, it is also possible to compare the performance of our analysis to other analysis refining \emph{may}/\emph{must} analysis, like persistence analysis or analysis performing virtual inlining and unrolling.

\bibliographystyle{abbrv}
\bibliography{JRWRTC2016}

\end{document}

%% file: unknownExample.tex
\begin{tikzpicture}[scale=0.8, every node/.style={scale=0.8}]

	\node[draw, rectangle, minimum size=0.8cm] (b1) at (0, 5) {1. $a$};
	\node[draw, rectangle, minimum size=0.8cm] (b2) at (0.6, 4) {2. $b$};
	\node[draw, rectangle, minimum size=0.8cm] (b3) at (0.6, 3) {3. $c$};
	\node[draw, rectangle, minimum size=0.8cm] (b4) at (0.6, 2) {4. $d$};
	\node[draw, rectangle, minimum size=0.8cm] (b5) at (0, 1) {5. $b$};
	\node[draw, rectangle, minimum size=0.8cm] (b6) at (0, 0) {6. $a$};

	\draw[->] (b1) to (b2);
	\draw[->] (b2) to (b3);
	\draw[->] (b3) to (b4);
	\draw[->] (b4) to (b5);
	\draw[->] (b5) to (b6);
	\draw[->] (b1) to (b5);

	\node[right of = b1, node distance = 0.4cm, anchor = west, align = left] {$[a,\bot,\bot,\bot]_{May}$ \\ $[a,\bot,\bot,\bot]_{Must}$};
	\node[right of = b2, node distance = 0.4cm, anchor = west, align = left] {$[b,a,\bot,\bot]_{May}$ \\ $[b,a,\bot,\bot]_{Must}$};
	\node[right of = b3, node distance = 0.4cm, anchor = west, align = left] {$[c,b,a,\bot]_{May}$ \\ $[c,b,a,\bot]_{Must}$};
	\node[right of = b4, node distance = 0.4cm, anchor = west, align = left] {$[d,c,b,a]_{May}$ \\ $[d,c,b,a]_{Must}$};
	\node[right of = b5, node distance = 0.4cm, anchor = west, align = left] {$[b,\{a,d\},c,\bot]_{May}$ \\ $[b,\bot,\bot,\bot]_{Must}$};
	\node[right of = b6, node distance = 0.4cm, anchor = west, align = left] {$[a,b,d,c]_{May}$ \\ $[a,b,\bot,\bot]_{Must}$};

\end{tikzpicture}

%% file: simplified_graph.tex
\tikzset{
	evict/.style={
		draw,
		isosceles triangle,
		isosceles triangle apex angle=60,
		shape border rotate=90
	},
	load/.style={
		draw,
		isosceles triangle,
		isosceles triangle apex angle=60,
		shape border rotate=-90
	},
}

\begin{tikzpicture}[scale=0.7, every node/.style={scale=0.8}]

	\node[load] (load1) {};
	\node[below right of = load1, node distance = 0.8cm] (zone1) {$[\dots]_a$};
	\node[evict, below left of = zone1, node distance = 0.8cm] (evict1) {};
	\path[->,draw] (load1) to (evict1);
	\path[->,draw] (load1) to[out=-70, in=70] (evict1);
	\path[->,draw] (load1) to[out=-110, in=110] (evict1);

	\node[below right of = evict1] (emptyzone) {$\varepsilon_a$};

	\node[load, below left of = emptyzone] (load2) {};
	\node[below right of = load2, node distance = 0.8cm] (zone2) {$[\dots]_a$};
	\node[evict, below left of = zone2, node distance = 0.8cm] (evict2) {};
	\path[->,draw] (load2) to (evict2);
	\path[->,draw] (load2) to[out=-70, in=70] (evict2);
	\path[->,draw] (load2) to[out=-110, in=110] (evict2);

	\path[->,draw, dashed] (evict1) to (load2);
	\path[->,draw, dashed] (evict1) to[out=-60, in=60] (load2);
	\path[->,draw, dashed] (evict1) to[out=-120, in=120] (load2);

	\node[load, below right of = load1, label = right:access to $a$, node distance = 2cm] (loadLegend) {};
	\node[evict, below of = loadLegend, label = right:eviction of $a$] (evictLegend) {};
\end{tikzpicture}

%% file: workflow.tex
\begin{tikzpicture}[scale=0.65, every node/.style={scale=0.65}]

\node[draw, dashed, ellipse, minimum width = 2cm, minimum height = 1cm] (cfg) at (4, 3) {CFG};
\node[draw, dashed, ellipse, minimum width = 2cm, minimum height = 1cm, align = center] (cache_configuration) at  (0, 3) {Cache\\Configuration};
\node[draw, dashed, ellipse, align=center, minimum height=1cm, minimum width=2cm] (classification) at (4, 0) {Hit/Miss\\Classification};

\node[draw, rectangle, align=center, minimum height=1cm, minimum width=2cm] (may_must) at (0, 1.5) {May/Must\\Analysis};
\node[draw, rectangle, align=center, minimum height=1cm, minimum width=2cm] (cache_modeling) at (4, 1.5) {Cache\\Modeling};
\node[draw, rectangle, align=center, minimum height=1cm, minimum width=2cm] (model_checker) at (8, 1.5) {Model\\Checker};

\draw[->] (cfg) -- (may_must);
\draw[->] (cache_configuration) -- (may_must);
\draw[->] (may_must) -- (cache_modeling) node[midway,above] {Unknown};
\draw[->] (may_must) -- (classification) node[midway,above, sloped] {Not unknown};
\draw[->] (cache_modeling) -- (model_checker);
\draw[->] (model_checker) -- (classification);
\draw[->] (cache_configuration) -- (cache_modeling);
\draw[->] (cfg) -- (model_checker);

\end{tikzpicture}